\documentclass[12pt]{iopart}
\usepackage{graphicx}
\usepackage{bm}
\begin{document}
\title{Acoustic resonances in two dimensional radial sonic crystals shells}
\author{Daniel Torrent and Jos\'e S\'anchez-Dehesa}
\ead{jsdehesa@upvnet.upv.es}
\address{Wave Phenomena Group, Departamento de Ingenier\'{\i}a Electr\'onica, Universidad Polit\'ecnica de Valencia, C/Camino de Vera s.n., E-46022 Valencia, Spain}%
\date{\today}
\begin{abstract}
Radial sonic crystals (RSC) are fluidlike structures infinitely periodic along the radial direction. They have been recently introduced and are only possible thanks to the anisotropy of specially designed acoustic metamaterials [see Phys. Rev. Lett. {\bf 103} 064301 (2009)]. We present here a comprehensive analysis of two-dimensional RSC shells, which consist of a cavity defect centered at the origin of the crystal and a finite thickness crystal shell surrounded by a fluidlike background. We develop analytic expressions demonstrating that, like for other type of crystals (photonic or phononic) with defects, these shells contain Fabry-Perot like resonances and strongly localized modes. The results are completely general and can be extended to three dimensional acoustic structures and to their photonic counterparts, the radial photonic crystals. 
\end{abstract}
\maketitle
%
%
\section{Introduction}
Acoustic metamaterials are a new type of structures with exciting properties. They consist of periodic arrangements of sonic scatterers embedded in a fluid or a gas and their extraordinary properties appear at wavelengths much larger than the lattice separation. Very recently, these authors employed acoustic metamaterials to demonstrate that is possible to create a new type of sonic crystals named Radial Sonic Crystals (RSC)\cite{torrentPRL09}.
 These crystals are radially periodic structures that, like the crystals with periodicity along the Cartesian axis, verified the Bloch theorem and consequently, the sound propagation is only allowed for certain frequency bands. Some interesting applications of RSC have been envisaged. For example, they are proposed to fabricate dynamically orientable antennas or acoustic devices for beam forming\cite{torrentPRL09}.

In this paper we present a comprehensive analysis of RSC shells, which are structures consisting of a central cavity surrounded by finite RSC, both being embedded in a fluid or a gas.  It will be shown that these structures generate two types of resonant modes; ones are strongly localized in the cavity the the others are weakly localized in the surrounding shell. Particularly, those strongly localized or cavity modes can have extremely large quality factors and made them suitable for developing acoustic devices in which localization is specially needed; for example, it can be used to enhance the phonon-photon interaction in mixed phononic-photonic structures. Moreover, they can be employed as acoustic resonators of high $Q$ (quality factor) since it is shown here that acoustic cavities based on RSC shells can achieve Q-values much larger than that reported up to date\cite{miklosRSI01}.   
  
The paper is organized as follows. After this introduction, in Section 2, the theory behind the infinitely periodic RSC is briefly introduced for the comprehensiveness of the work. Section 3 studies the resonance properties of a particular RSC shell and special emphasis is put in getting analytical formulas that give a physical insight of the properties associated to the two type of resonant modes. The mode pressure patterns are analyzed in Section 4. Finally, the work is summarized in Section \ref{sec:summary}, where we also give a perspective of possible application of these structures as well as an account of future work to be performed. 

%
%
\section{\label{sec:introduction}Radial Sonic Crystals}
The propagation of a harmonic acoustic wave, $P({\bf r};\omega)$, in an anisotropic fluid is given by the following wave equation \cite{torrentPRB09}:
\begin{equation}
B\bm{\nabla}\left(\rho^{-1}\cdot\bm{\nabla}\right)P+\omega^2P=0,
\end{equation} 
where $B$ and $\rho^{-1}$ are the bulk modulus and the reciprocal of the mass density tensor, respectively. 

By assuming that the propagation takes place in 2D, the pressure field $P$ can be factorized as follows,
\begin{equation}
P(r,\theta;\omega )=\sum_qP_q(r)e^{iq\theta},
\end{equation}
where $(r,\theta)$ are the cylindrical coordinates and $P_q(r)$ accomplishes the equation
\begin{equation}
\label{eq:waveqr}
 \frac{\partial}{\partial r}\left(\frac{r}{\rho_r}\frac{\partial}{\partial r}\right)P_q(r)+\left[\frac{r}{B}\omega^2-\frac{q^2}{r\rho_\theta}\right]P_q(r)=0,
\end{equation}
where $q$ are discrete number ($q=$0,1,2,...)$\rho_r$ and $\rho_\theta$ being, respectively, the components of the anisotropic mass density tensor in 2D.

Due to anisotropy, all the coefficients in \eref{eq:waveqr} are independent and, therefore, equation \eref{eq:waveqr} can be made invariant under the set of translations $r\to r+nd$, where $n$ is an integer and $d$ is the period, which is prefixed parameter. 
Acoustic structures having these properties are called Radial Sonic Crystal (RSC)\cite{torrentPRL09} because they are formally equivalent to the crystals already known, like atomic crystals, photonic crystals or phononic crystals, whose corresponding wave equations are invariant under translations along the Cartesian axis. 
\par
\subsection{Acoustic parameters radially periodic}
\label{sec:rwc}
The conditions that leave \eref{eq:waveqr} invariant under translations of the type $r\to r+nd$ are:
\numparts
\label{eq:coeff}
\begin{eqnarray}
 \frac{r+nd}{\rho_r(r+nd)}&=&\frac{r}{\rho_r(r)}\\
\frac{r+nd}{B(r+nd)}&=&\frac{r}{B(r)}\\
(r+nd)\rho_\theta(r+nd)&=&r\rho_\theta(r)
\end{eqnarray}
\endnumparts
From these conditions is easily concluded that the acoustic parameters can be cast into the form\cite{torrentPRL09}: 
\begin{figure}
\centering
\includegraphics[scale=0.8]{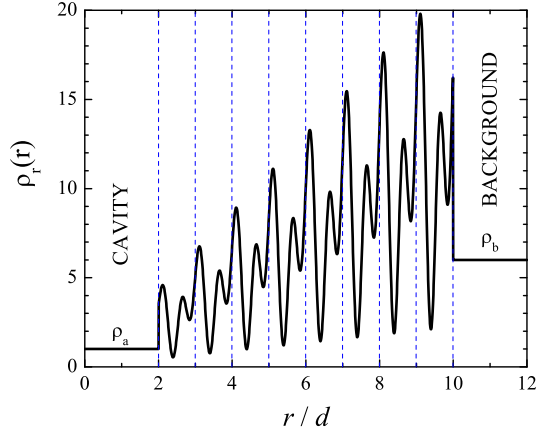} 
\caption{\label{fig:radial_parameter}Plot of a radially periodic parameter. It is seen how there is a divergence at the origin, but in practice the finite structure will consist in a cavity region, the crystal and the background.}
\end{figure}
\par
\numparts
\label{aparameters}
\begin{eqnarray}
 \rho_r(r)&=&r\hat{\rho}_r(r)\\
\rho^{-1}_\theta(r)&=&r\hat{\rho}_\theta^{-1}(r)\\
B(r)&=&r\hat{B}(r),
\end{eqnarray}
\endnumparts
where $\hat{\rho_r}(r), \hat{\rho_\theta}(r)$ and $\hat{B}(r)$ represent periodic functions of $r$. Let us remark that any periodic function is valid to get the invariance obtained from (4). 

Figure \ref{fig:radial_parameter} depict, as a typical example, the profile of $\rho_r(r)$, which has been generated by the product of $r$ with a radially periodic function $\hat{\rho_r}$, as explained in (5a). Notice that the RSC is only defined in the region 2$\leq r/d \leq$ 10. So, this plot also illustrates the case of an acoustic cavity generated by a RSC shell of 8 layers; below $r/d=$2 the mass density is $\rho_a$ and for $r/d>$10 the surrounded background has density $\rho_b$.
\subsection{Wave Propagation and Band Structure}	
\label{sec:bandstructure}
The periodicity of coefficients \eref{eq:coeff} allows to apply the Bloch's theorem to solve \eref{eq:waveqr}. Thus, the solutions $P_q(r)$ can be chosen to have the form of a plane wave times a function with the periodicity of the lattice\cite{Ascroft76}:
\begin{equation}
\label{ec:pq}
P_q(r)=e^{ik_qr}\sum_nP_{qn}e^{iG_nr}
\end{equation} 
where $k_q$ is the wavenumber and $G_n$ are the reciprocal lattice vectors given by $G_n=2n\pi/d$, $n$ being integers.
\par
The periodicity of coefficients allows the expansion in Fourier series:
\numparts
\begin{eqnarray}
 \frac{r}{\rho_r}=\sum_nf_ne^{iG_nr}\\
\frac{1}{r\rho_\theta}=\sum_ng_ne^{iG_nr}\\
\frac{r}{B}=\sum_nh_ne^{iG_nr}
\end{eqnarray} 
\endnumparts
The insertion of these expressions in \eref{eq:waveqr} and after using the expansion of the field \eref{ec:pq}, the following matrix equation is found
\begin{equation}
\bm{M}\bm{P}=\omega^2\bm{N}\bm{P},\\ 
\end{equation}
where the matrix elements are
\numparts
\begin{eqnarray}
M_{mn}&=&f_{m-n}(K_q+G_m)(K_q+G_n)+q^2g_{m-n}\\
N_{mn}&=&h_{m-n}
\end{eqnarray} 
\endnumparts
\begin{figure}
\centering
\includegraphics[]{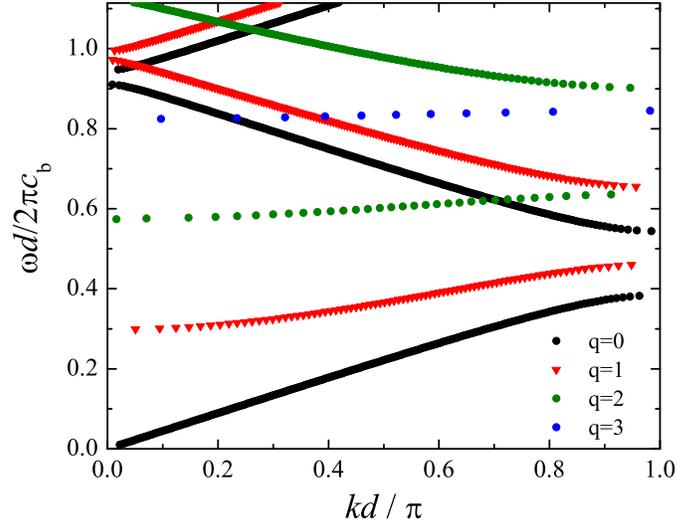} 
\caption{Acoustic band structure for a two dimensional radial sonic crystal.}
\label{fig:bands}
\end{figure}
\par
This matrix equation is a generalized eigenvalue problem that can be solved numerically with a wide variety of methods. Thus, once the acoustic parameters defining the crystal are given, we can solve the eigenvalue equation $\omega=\omega(K_q)$, obtaining a dispersion relation that is different for each multipole $q$ and noting that $\omega(K_q)=\omega(K_{-q})$\cite{Ascroft76}. 

The final form for the field inside the crystal is given by a linear combination of plane waves. 
\begin{equation}
 P(r,\theta)=\sum_qe^{iK_qr}e^{iq\theta}\sum_nP_{qn}e^{iG_n}
\end{equation}
which is a linear combination of plane waves. 
\par

In \cite{torrentPRL09}, we have considered a RSC made of two types of alternating ``homogeneous'' layers. Actually, the layer are not exactly ``homogeneous'' because they need to be radially dependent. The term ``homogeneous'' applies here because this medium is the equivalent to the system ``ABABABA...'' employed in standard heterostructures. The difference is that in this case $\hat{\rho}_r,\hat{\rho_\theta}$ and $\hat{B}$ are considered as the ``stepped'' functions.
\subsection{Low Frequency Bandgaps}
Figure \ref{fig:bands} shows that in the low frequency limit only the dispersion relation corresponding to modes $q=$0 (black lines), has a linear behavior near the zero frequency. 
The bands corresponding to higher modes, $q>$0 present low frequency bandgaps. In other words, for non monopolar modes, the RSC behaves like a ``radially homogeneous'' medium; that is, a system where the periodic functions $\hat{\rho}_r$, $\hat{\rho}_\theta$ and $\hat{B}$ take constant values $\overline{\rho}_r$, $\overline{\rho_\theta}$ and $\overline{B}$, respectively. 
Then, in such a medium the wave equation \eref{eq:waveqr} becomes
\begin{equation}
\frac{\partial^2P_q(r)}{\partial
r^2}+\left[\omega^2\frac{\overline{\rho}_r}{\overline{B}}-q^2\frac{\overline{\rho}_r}{\overline{\rho_\theta}}\right]P_q(r)=0
\end{equation}
The solutions of this equation are plane waves
\begin{equation}
P_q(r)=e^{ik_qr},
\end{equation}
where
\begin{equation}
k_q^2=\omega^2\frac{\overline{\rho}_r}{\overline{B}}-q^2\frac{\overline{\rho}_r}{\overline{\rho_\theta}};\ \ \ q=0,1,2,\ldots
\end{equation}
Therefore, below certain cut-off frequencies given by
\begin{equation}
\omega_{q_c}=q\sqrt{\frac{\overline{B}}{\overline{\rho}_\theta}}
\end{equation}
the wavenumbers take imaginary values and evanescent waves are the only possible solutions to \eref{eq:waveqr}. 
The cut-off frequency defines the low-frequency bandgaps of the corresponding mode and it is a property characterizing the so called ``radially homogeneous'' medium, whose acoustic parameters (density and bulk modulus) linearly increases with the radial distance to the origin of the crystal.
%
%
\section{Resonances in a RSC shell}
\label{sec:RSCshell}
Like with other type of crystals, it is not possible to handle infinite RSC. 
So, let us consider a RSC shell consisting of 8 periods as the one described in Fig. \ref{fig:radial_parameter}. 
The shell has an inner cavity of radius $R_a=2d$ that acts like a {\it defect}. 
A scheme of the structure under study is shown in figure \ref{fig:radial_parameter}, where it can be seen that the cavity medium $A$ is enclosed by the anisotropic shell $S$ and both being surrounded by a background fluid $B$ that exists for $r\geq R_b$. 
Both the medium $A$ and $B$ are homogeneous and isotropic fluidlike materials with acoustic parameters $\rho_a$, $B_a$ and $\rho_b$, $B_b$, respectively. 
The shell is assumed to be an anisotropic fluid-like material satisfying \ref{eq:waveqr}.

Here, we develop a formalism similar to the transfer matrix method \cite{bendicksonPRE96} to study two type of scattering problems. 
First, we consider the scattering of an external sound wave impinging the RSC shell.
Afterwards, we analyze the case when the exciting sound is inside the cavity shell.
\par
In both media, $A$ and $B$, the fields are given by a linear combination of Hankel and Bessel functions,
\begin{equation}
P(r,\theta)=\sum_q\left[C_{\ell q}^+H_q(k_\ell r)+C_{\ell q}^-J_q(k_\ell r)\right]e^{iq\theta},
\end{equation}
where $k_\ell=\omega/c_\ell$ and $\ell=a,b$.
\par

For the anisotropic shell $S$, which covers the region $R_a\leq r \leq R_b$, the pressure field can be expressed as 
\begin{equation}
 P(r,\theta)=\sum_q\left[C_{sq}^+\phi_q^+(r)+C_{sq}^-\phi_q^-(r)\right]e^{iq\theta},
\end{equation} 
where $\phi_q^+$ and $\phi_q^+$ are two linearly independent solutions of the wave equation. 
The proposed solutions in the shell are completely general and it is assumed that the acoustic parameters $\rho_r(r),\rho_\theta(r)$ and $B(r)$ are arbitrary functions of the radial coordinate $r$. Later we will consider a particular RSC. 
\par
\par 
To determine the relationships between coefficients $C_{\ell q}^\pm$ for $\ell=a,b,s$ boundary conditions must be applied at the boundaries $R_a$ and $R_b$. 
The boundary conditions are: (1)the continuity of pressure and (2) the continuity of the radial component of particle velocity. 
Because of the radial symmetry of the problem it is possible to uncouple the equations for the different multipole components $q$.

Therefore, at $r=R_\ell$, for $\ell=a$, $b$, we require that:
\numparts
\begin{eqnarray}
\fl C_{\ell q}^+H_q(k_{\ell}R_{\ell})+C_{\ell q}^-J_q(k_{\ell} R_{\ell})&=&C_{sq}^+\phi_+(R_{\ell})+C_{sq}^-\phi_-(R_{\ell})\\
\fl \frac{k_\ell R_\ell}{\rho_l}\left[C_{\ell q}^+H'_q(k_\ell R_\ell)+C_{\ell q}^-J'_q(k_lR_l)\right]&=& \frac{R_\ell}{\rho_r(R_\ell)}\left[C_{sq}^+\partial_r\phi_+(r)+C_{sq}^-\partial_r\phi_-(r)\right]_{r=R_\ell}
\end{eqnarray}
\endnumparts
Note that, for convenience, the continuity equation of particle velocity has been multiply by the radial coordinate $R_{\ell}$ at both sides.

\par 
Solving for the coefficients $C_{sq}^\pm$ the transfer matrix {\bf M} of the shell is found
\begin{equation}
\left(\begin{array}{c}C_{aq}^+\\C_{aq}^-\end{array}\right)=
M
\left(\begin{array}{c}C_{bq}^+\\C_{bq}^-\end{array}\right)
\end{equation} 
After some algebra, the matrix {\bf M} can be factorized as ${\bf M}={\bf M}_a^{-1}{\bf M}_s{\bf M}_b$, where,
\numparts
\begin{eqnarray}
{\bf M}_a^{-1}&=&\frac{i\pi \rho_a}{2}\left(\begin{array}{cc}k_aR_a\rho_a^{-1}J'_q(k_aR_a) & -J_q(k_aR_a)\\ -k_aR_a\rho_a^{-1}H'_q(k_aR_a) & H_q(k_aR_a)\end{array}\right)\\
{\bf M}_s&=&\left(\begin{array}{cc}\phi_{11} & R_b^{-1}\rho_r(R_b)\phi_{12}\\ R_a\rho_r^{-1}(R_a)\phi_{21} & R_b^{-1}R_a\rho_r(R_b)\rho_r^{-1}(R_a)\phi_{22}\end{array}\right)\\
{\bf M}_b&=&\left(\begin{array}{cc} H_q(k_bR_b) & J_q(k_bR_b)\\k_bR_b\rho_b^{-1} H'_q(k_bR_b) & k_bR_b\rho_b^{-1}J'_q(k_bR_b)\end{array}\right)
\end{eqnarray} 
\endnumparts
\par
The quantities $\phi_{ij}$ in ${\bf M}_s$ are Wronskians of the shell function $\phi_q^\pm(r)$ and their derivatives, so that no analytical simplification can be done here. 
However it is easy tho show that the determinant of ${\bf M}_s$ is one (see the Appendix). 
\par
Now, the coefficients for the two type of scattering problems of interest here can be obtained by applying the transfer matrix formalism.  

First, let us consider the case when there is no sound source inside the the cavity and, therefore, $C_{aq}^+=0$.
The exciting field is an external wave defined by $C_{bq}^-=A_q^0$, and the goal is to find the coefficients $C_{aq}^-\equiv T_q^iA_q^0$ and $C_{bq}^+\equiv T_q^{sc}A_q^0$, by solving
\begin{equation}
\left(\begin{array}{c} \bm{0}\\T_q^{in}\end{array}\right)=
\left(\begin{array}{cc} M_{11} & M_{12} \\ M_{21} & M_{22} \end{array}\right)
\left(\begin{array}{c} T_q^{sc} \\ \bm{1}\end{array}\right)
\end{equation} 
That is,
\numparts
\begin{eqnarray}
 T_q^{sc}&=&-\frac{M_{12}}{M_{11}}\\
 T_q^{in}&=&\frac{M_{22}M_{11}-M_{12}M_{21}}{M_{11}},
\end{eqnarray}
\endnumparts
where the numerator of $T_q^{in}$ is the determinant of ${\bf M}$, which is $|{\bf M}|=|{\bf M}_a||{\bf M}_s||{\bf M}_b|=\rho_a/\rho_b$. This yields
\numparts
\begin{eqnarray}
 T_q^{sc}&=&-\frac{M_{12}}{M_{11}}\\
 T_q^{in}&=&\frac{\rho_a}{\rho_b}\frac{1}{M_{11}}
\end{eqnarray}
\endnumparts

If the exciting field is put inside the cavity, the pressure wave is defined by the coefficients $C_{aq}^+= A_q^0$. Now, a standing wave is excited inside the cavity $C_{aq}^-\equiv R_qA_q^0$ and
an acoustic field defined by $C_{bq}^+\equiv T_qA_q^0$ and $C_{bq}^-=0$ is radiated by the cavity. 
This problem is solved by obtaining the coefficients $R_q$ and $T_q$ in the following matrix equation:
\begin{equation}
\left(\begin{array}{c} \bm{1}\\R_q \end{array}\right)=
\left(\begin{array}{cc} M_{11} & M_{12} \\ M_{21} & M_{22} \end{array}\right)
\left(\begin{array}{c} T_q \\ \bm{0}\end{array}\right)
\end{equation} 
Again, solving for the coefficients,
\numparts
\begin{eqnarray}
 R_q&=&-\frac{M_{21}}{M_{11}}\\
 T_q&=&\frac{1}{M_{11}}
\label{eq:tq}
\end{eqnarray}
\endnumparts

For the two scattering problems analyzed above, the behavior of their corresponding coefficients is controlled by a common denominator; the diagonal element $M_{11}$. Moreover, $T_q^{in}=T_q$ when both the cavity and the background are the same medium; i.e., $\rho_a=\rho_b$. 

If the shell is made of a RSC with $N$-periods, the matrix ${\bf M}_s$ of each period is identical, due to the periodicity defining the RSC. 
The transfer matrix of such system is therefore
\begin{equation}
{\bf M}={\bf M}_a^{-1}{\bf M}_s^N{\bf M}_b
\end{equation} 
\par
Since the ${\bf M}_s$ matrix has modulus one, its eigenvalues are of the form $\lambda_\pm=e^{\pm iK_qd}$, being $K_q=K_q(\omega)$ the dispersion relation that defines the band structure of the crystal. 
These properties make possible express the $N$-th power of the matrix $M_s$ as\cite{bendicksonPRE96}
\begin{equation}
\label{eq:msN}
 M_s^{N}=M_s\frac{\sin NK_qd}{\sin K_qd}-I\frac{\sin (N-1)K_qd}{\sin K_qd},
\end{equation} 
\label{eq:mN}
and the transfer matrix of the system as
\begin{equation}
\label{eq:transmat}
 M=M_a^{-1}M_sM_b\frac{\sin NK_qd}{\sin K_qd}-M_a^{-1}M_b\frac{\sin (N-1)K_qd}{\sin K_qd}
\end{equation} 
 The complex frequencies $\hat{\omega}$ solving equation $M_{11}=0$ completely characterize the resonant modes in the structure; i.e., their frequencies and lifetimes. 
 However, the numerical solution of such equation is a cumbersome task and it is beyond the scope of the present work. 
 Here, we prefer to give a physical insight of the resonances $\omega$ for which the coefficients have large or small values.
\par
Two types of resonances can be distinguished: cavity modes and Fabry-Perot like resonances. 
The former are strongly localized inside the cavity shell, the later are due to the finite thickness of the RSC making the shell.
\subsection{\label{subsec:cavity}Cavity modes}
These resonances occur at frequencies within the bandgaps of the RSC. 
They are characterized by their strong localization inside the cavity shell as it is shown below. 

Bloch wavenumbers within the bandgaps are complex number, $K_q\approx i\hat{K}_q$, and as a consequence the periodic functions of $NK_qd$ in \eref{eq:mN} become hyperbolic functions of $N\hat{K}_qd$. Therefore, the dominant term in the expression for $M_{11}$ are proportional to $\sin NK_qd\approx \frac{1}{2}e^{N\hat{K}_qd}$. 
If we consider, for example, that the sound source is placed inside the cavity shell, the field radiated out of the shell is given by (24b), that is
\begin{equation}
T_q=1/M_{11}\approx e^{-N\hat{K}_qd},
\end{equation} 

This exponentially decreasing behavior with increasing $N$ has been previously characterized in other type of crystals, like photonic or phononic crystals. For example,  the attenuation by a phononic crystal slab has been demonstrate to be exponentially decreasing with the number of crystal periods\cite{goffauxAPL03}.
\par
\begin{figure}
\centering
\includegraphics[]{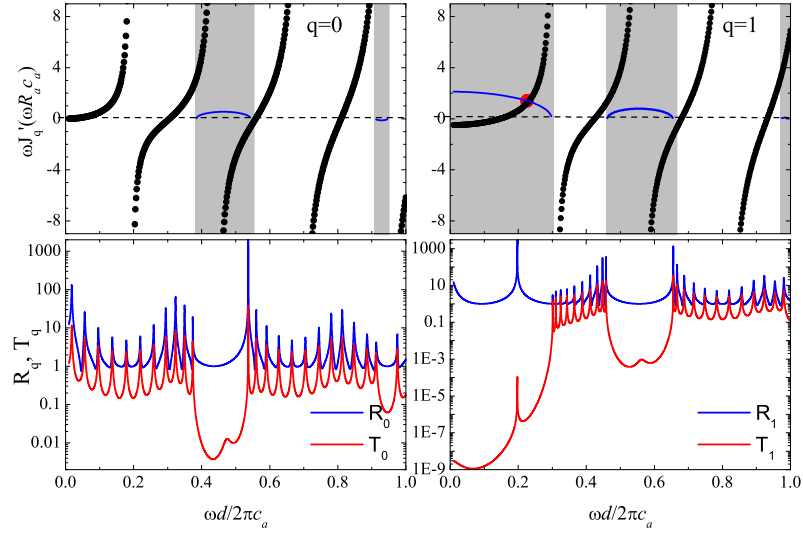} 
\caption{\label{fig:modosq01}Localized modes and Fabry-Perot resonances for a finite RSC consisting of 8 periods and an inner cavity of radius $R_a=2d$ (see Fig. 1). Top panels: Graphic resolution of Eq. (20) for modes 0 and 1, respectively. The dots define the frequency positions at which the mode cavities are predicted. The shadow zone define the acoustic bandgaps shown in Fig. 2. Bottom panels: The corresponding scattering coefficients, $T_0$ and $T_1$, which have peaks at the frequencies approximately determined by Eq. (20) (see text).}
\end{figure}
\begin{figure}
\centering
\includegraphics[]{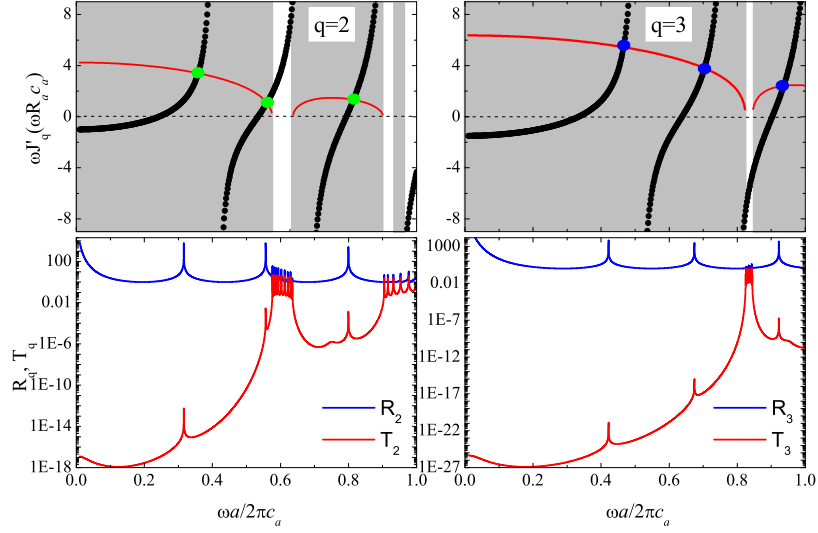} 
\caption{\label{fig:modosq23}The same representation than in Fig. \ref{fig:modosq01} but for modes $q=$2 and $q=$3, respectively. Notice that the large frequency bandgaps (shadow regions) appearing at low frequencies (see also Fig. \ref{fig:bands} The Fabry-Perot like resonances are shown in the narrow passband frequencies (white regions).}
\end{figure}
\par
However it is also possible to demonstrate that, in the bandgap region, $T_q$ have strong transmission peaks, which are associated to the existence of cavity modes. Thus, after some algebra, the $M_{11}$ element can be cast as
\begin{equation}
 M_{11}=\frac{i\pi \rho_a}{2}\left[X_qH_q(k_bR_b)+Y_qk_bR_b\rho_b^{-1}H_q'(k_bR_b)\right]
\end{equation} 
where
\numparts
\begin{eqnarray}
\label{eq:Xqeq0}
 X_q&=&M_{11}^N k_a\rho_a^{-1}R_aJ_q'(k_aR_a)-J_q(k_aR_a)M_{21}^N\\
\noindent \rm{and}\nonumber\\
 Y_q&=&M_{12}^N k_a\rho_a^{-1}R_aJ_q'(k_aR_a)-J_q(k_aR_a)M_{22}^N
\end{eqnarray}
\endnumparts 
When $X_q=0$ we have that
\begin{equation}
 Y_q=-J_q(k_aR_a)/M_{11}^N\approx e^{-N\hat{K}_qd},
\end{equation} 
which means that now $M_{11}$ behaves inversely as before; i.e., the transmission coefficient increases exponentially with the number $N$ of layers.
\begin{equation}
T_q=1/M_{11}\approx e^{N\hat{K}_qd}
\end{equation} 
A similar behavior is found after analyzing the case $Y_q=0$.
\par
The resonance frequency can be obtained from the condition $X_q=0$. The following condition must be hold 
\begin{equation}
 k_a\rho_a^{-1}R_a\frac{J_q'(k_aR_a)}{J_q(k_aR_a)}=\frac{M_{21}^N}{M_{11}^N }\approx R_a\rho_r^{-1}(R_a)\frac{\phi_{21}}{\phi_{11}},
\end{equation} 
where the quantities $\phi_{ij}$ have been already introduced as factors in the elements of ${\bf M}_s$ [see (19b)]. 
Since we are working at bangap frequencies, the ratio $\phi_{21}/\phi_{11}$ can be approximated by
\begin{equation}
 \frac{\phi_{21}}{\phi_{11}}\approx i K_q=-\textit{Im}[ K_q(\omega)]
\end{equation}  
Therefore, the equation for the frequency $\omega_0$ of the cavity modes is rapidly find
\begin{equation}
\label{eq:condXq0}
\omega_0 J_q'(\omega_0 R_a/c_a)=-\frac{\rho_ac_a}{\rho_r(R_a)}\textit{Im}[K_q(\omega_0)]J_q(\omega_0 R_a/c_a),
\end{equation}
which can be solved graphically. 

Figures \ref{fig:modosq01} and \ref{fig:modosq23} show the graphical solution of \eref{eq:condXq0}, where the left hand side is depicted with black dots symbols while the red lines represent the right hand side. The color dots at the crossing points give the (approximate) values of the cavity modes. 
These values are confirmed by the numerical solution of coefficients $R_q$ and $T_q$, which are depicted in the bottom panels of Figs. \ref{fig:modosq01} and \ref{fig:modosq23}. Note how, at frequencies within the bandgaps (shadowed zones in the top panels), the coefficients show peaks that practically coincide with the points giving the solutions of \eref{eq:condXq0}. 
The peaks in the passband regions (white zones) correspond to Fabry-Perot like modes that are discussed in the following subsection. 
Let us remark that coefficients $R_q$ and $T_q$ in Figs. \ref{fig:modosq01} and \ref{fig:modosq23} have been calculated with a resolution of $10^{-5}$ along the frequency axis and, therefore, their values at the peaks are not necessarily converged. For an exact determination of peak heights lower frequency steps should be chosen till the convergence is obtained. 

\begin{figure}
\centering
\includegraphics[]{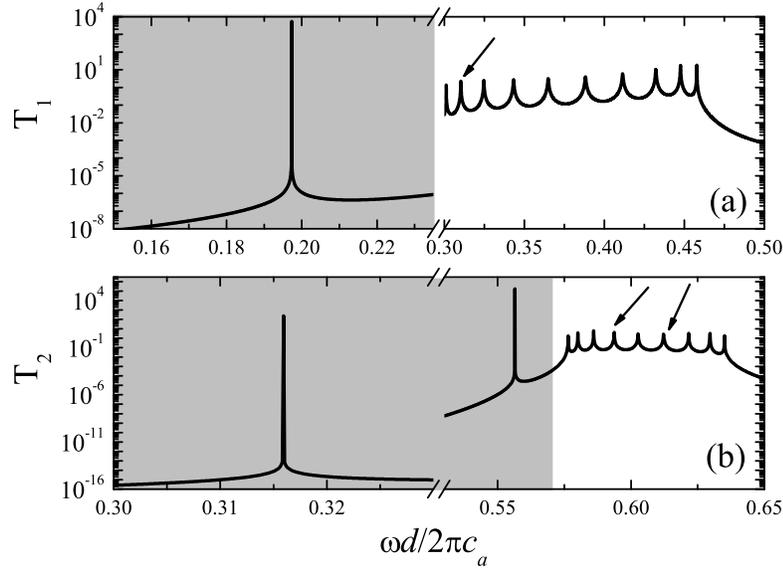} 
\caption{\label{fig:Zoom}(a) Coefficients $T_1$ of a RSC shell with the same parameters than in Fig. \ref{fig:modosq01}. Nut now the calculation is performed with frequency steps of $10^{-7}$ units. (b) Coefficients $T_2$ of a RSC shell with the same parameters than in Fig. \ref{fig:modosq23}. But now the calculation has been performed with frequency steps of $10^{-12}$ units. The shadowed regions in both figures define the low frequency bandgaps of the dispersion relations with symmetry $q=$1 and $q=$2, respectively, in the corresponding RSC (see Fig. \ref{fig:bands}).}
\end{figure}

The behavior of coefficients $T_q$ near resonances suggests that their moduli have the following functional form
\begin{equation}
 |T_q(\omega)|^2=\frac{1}{A(\omega-\omega_0)^2e^{N\hat{K}_qd}+Be^{-N\hat{K}_qd}}
\end{equation} 
The quality factor $Q=\omega_0/\Delta \omega$, where $\Delta \omega$ is the full width at half maximum (FWWHM) of the resonance profile. In other words, $\Delta \omega=\omega_+-\omega_-$, where $\omega_{\pm}$ being the frequencies at which $|T_q(\omega_{\pm})|^2=|T_q(\omega_0)|^2/2$. It is easy to show that
\begin{equation}
 \omega_{\pm}=\omega_0\pm\sqrt{\frac{B}{A}},
\end{equation} 
and therefore
\begin{equation}
\label{eq:qfactorN}
 Q=2\sqrt{\frac{A}{B}}e^{N\hat{K}_qd}
\end{equation} 
\begin{figure}
\centering
\includegraphics[]{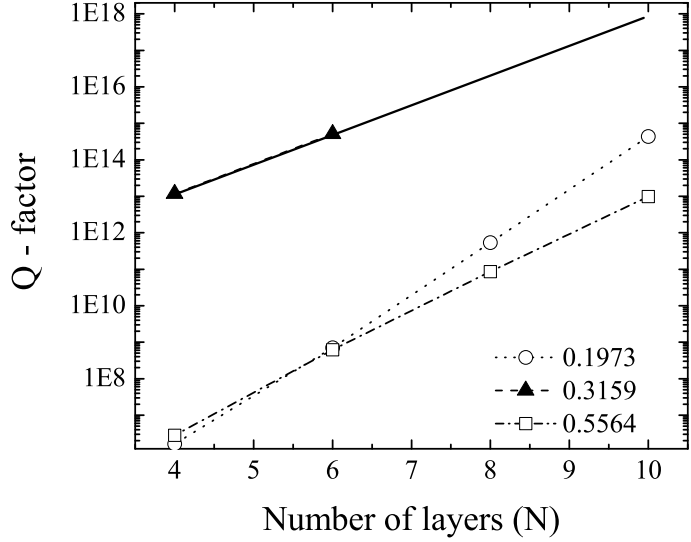} 
\caption{\label{fig:qfactors}The quality factor of three different modes localized in the cavity of a RSC shell made of N periods. The mode with frequency 0.1973 has dipolar symmetry ($q=$1) while modes with frequencies 0.3167 and 0.5538 both have quadrupolar symmetry ($q=$2).}
\end{figure}

This formula predicts that the $Q-$factor of a cavity mode grows up exponentially with the number $N$ of layers that form the RSC shell. 
This result is also known in the field of photonic crystals, where it was numerically demonstrated, for example, that the $Q-$factor of a cylindrical defect surrounded by a finite metallic 2D photonic crystal exponentially increases with the number of layers surrounding the cavity defect\cite{ochiaiPRB02}.

To demonstrate the validity of \eref{eq:qfactorN} we have calculated $Q$ as a function of N for three different cavity modes. It must be pointed out that $Q$ is obtained directly from the peaks in $T_q$ and, therefore, a very accurate determination of peak's height at the resonant frequencies is needed. This requirement is accomplished by exploring the frequency axis in very small steps until convergence is confirmed. As an example, Fig. \ref{fig:Zoom} shows the calculation of $T_1$ and $T_2$, which were already depicted in \ref{fig:modosq01} and \ref{fig:modosq23}, by using frequency steps of $10^{-7}$ and $10^{-12}$, respectively. Now, $T_q$ have values at the frequency resonances much larger than in Figs. \ref{fig:modosq01} and \ref{fig:modosq23} since are converged. 

Figure \ref{fig:qfactors} represents the $Q-$factor calculated for several values of $N$. We have analyzed one dipolar mode and two quadrupolar modes. The dipolar mode, which has frequency 0.1973, corresponds to the resonant peak having the lowest frequency appearing in Fig. \ref{fig:Zoom}(a). The two quadrupolar modes have frequencies 0.3167 and 0.5538, respectively, and are shown in Fig.\ref{fig:Zoom}(b). It is observed that the predicted exponential dependence is confirmed by the numerical simulations and provide the order of magnitude of $Q$ for these new structures. 
These structures used as acoustic resonator could provide $Q$ values much larger than that already obtained by perfect cylinders or spherical resonators\cite{miklosRSI01} 
\subsection{Fabry-Perot Resonances}
Fabry-Perot like resonances are generated by the oscillating terms $\sin NK_qd$ and $\sin (N-1)K_qd$ in \eref{eq:mN}, where $Nd$ defines the RSC thickness, $Nd=R_b-R_a$. These resonances are confined inside the RSC region. 
They are equivalent to Fabry-Perot modes already characterized for sonic crystals slabs\cite{sanchisPRB03}.  
A typical feature of these resonances is that they are frequency equidistant in the regions where the dispersion relation is linear. 
This feature is clearly observed in the profile of coefficients $R_0$ and $T_0$ shown in Fig. \ref{fig:modosq01}, where the peaks in the lower part of the first frequency band are equally spaced.
\par
The Fabry-Perot frequency oscillations, which are produced by the periodic functions of $K_q(R_b-R_a)$ in \eref{eq:transmat}, are coupled with the oscillating behavior of the quasi-periodic functions with argument $k_aR_a$ embedded in ${\bf M}_a$. 
Therefore, two parametric regimes can be distinguished. 
First, the regime of thick shells occur when $R_b >> R_a$ and, consequently, resonances localized in the RSC are dominants and the cavity modes will be scarcely shown and will appear as additional peaks on top of the oscillating profile of $R_q$ and $T_q$. 
The second regime is define by the condition $R_b\approx R_a$ and correspond to the case of thin shells. In this case the oscillating profile will be dominate by resonant peaks associated to cavity modes and, therefore, Fabry-Perot oscillations, if any, will appear in between.


Fabry-Perot modes have $Q-$factors much lower than that of cavity modes. For example, for the mode with dipolar symmetry located at frequency 0.3649 [see the arrow in Fig. \ref{fig:Zoom}(a)] we get $Q=$380. The modes with quadrupolar symmetry located at 0.5936 and 0.6122 [see the arrows in Fig. \ref{fig:Zoom}(b)] have $Q$ values of 4151 and 3364, respectively. Lower $Q-$values means less localization and, consequently, easy excitation by external sources. At this point it is interesting to point out that the existence of a great number of resonances combined with their easy excitation could be employed to design acoustic devices for broadband detection of external sound sources as was suggested in \cite{torrentPRL09}.

\section{Modes pressure fields}
\label{sec:modefields}
\begin{figure}
\centering
\includegraphics[]{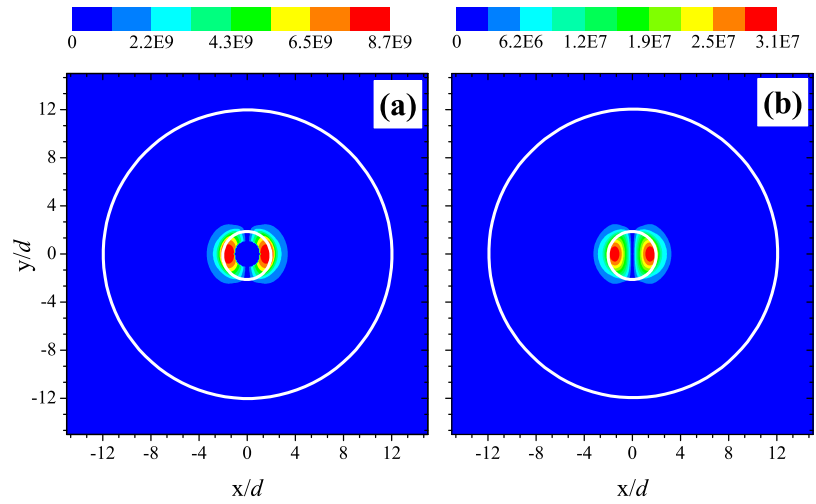} 
\caption{\label{fig:modecavityq1}Pressure maps (amplitude) of the cavity mode with frequency 0.1973 (in reduced units) and dipolar symmetry existing in the RSC shell made of 10 layers (see also Fig. \ref{fig:modosq01}). (a) Mode excited by a punctual sound source (not depicted) put at position $(x,y)=$(0,1) inside the cavity. (a) Mode excited by an external sound wave with a plane wavefront. The white circles define the borders of the RSC shell.}
\end{figure}
\begin{figure}
\centering
\includegraphics[]{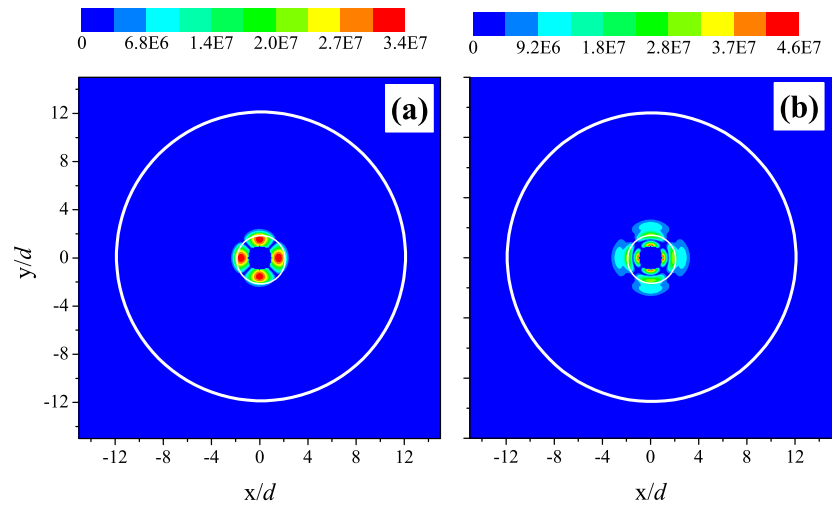} 
\caption{\label{fig:modecavityq2in}Pressure maps (amplitude) of the two cavity modes existing in the RSC shell with quadrupole symmetry ($q=$2). Their frequencies are 0.3159 (a) y 0.5564 (b) in reduced units. They are excited by a punctual sound source (not depicted) put at position $(x,y)=$(0,1) inside the cavity. The white circles define the borders of the RSC shell.}
\end{figure}
\begin{figure}
\centering
\includegraphics[]{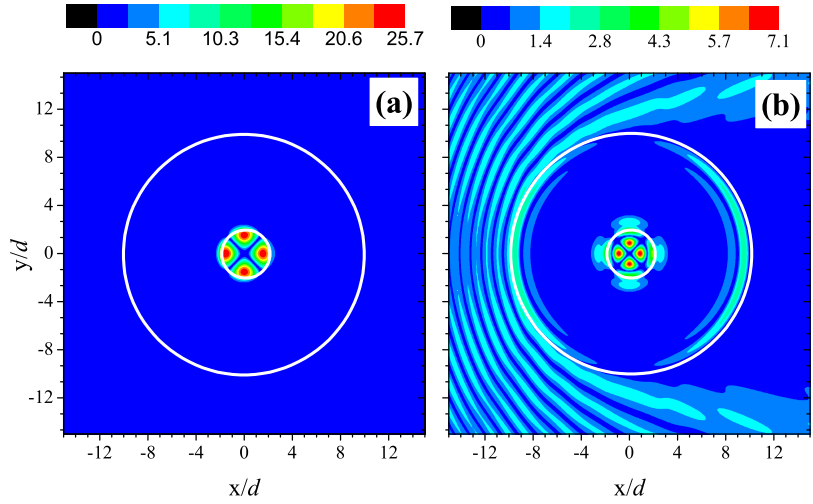} 
\caption{\label{fig:modescavityq2ext}Pressure maps (amplitude) calculated for the scattering of a sound plane wave impinging a RSC shell made of 8 layers. (a) External sound with frequency 0.3159 (in reduced units), which is resonant with the first quadrupole cavity mode in Fig. \ref{fig:Zoom}(b).
(b) External sound with frequency 0.5564 (in reduced units), which is resonant with the second quadrupole cavity mode in Fig. \ref{fig:Zoom}(b).}
\end{figure}
In this section we use pressure maps to discuss in brief the salient features of the different resonant modes that can be excited in a RSC shell. 
As typical examples we consider modes with dipole ($q=$1) and quadrupole ($q=$2) symmetry to illustrate their behavior under different excitation conditions. Cavity modes and Fabry-Perot like modes are shown. As it is pointed out before, the main feature distinguishing the cavity modes from the Fabry-Perot like modes is their difficulty of being excited by using external sound sources.

Figures \ref{fig:modecavityq1}(a) and \ref{fig:modecavityq1}(b) depict a cavity mode with dipole symmetry localized in a RSC shell made of 10 layers. The maps are obtained under the two different excitation conditions reported in Section 3: (a) by using a punctual sound source placed out of the cavity center and (b) by using an external sound with a plane wavefront. In both cases the exciting sound has the same frequency than the excited mode. Note the in both cases the cavity mode is excited and strong localization inside the cavity is clearly observed. 
The black circle in \ref{fig:modecavityq1}(a) represent the region not calculated by our computer code due to the presence of the exciting source.

 Figures \ref{fig:modecavityq2in} and \ref{fig:modescavityq2ext} represent the pressure maps of the two quadrupolar modes whose Q-factors were analyzed in Fig. \ref{fig:qfactors}. Figure \ref{fig:modecavityq2in} has been obtained by using an inner punctual source while Fig. \ref{fig:modescavityq2ext} was obtained by using the scattering of an external sound wave interacting resonantly with the cavity mode. In \ref{fig:modescavityq2ext} the thickness of the RSC shell is 8 layers since the modes are easier excited from outside. The oscillations observed in \ref{fig:modescavityq2ext}(b) out of the shell are produced by the external sound coming from the left side of the $x-$axis. Like the cavity mode with dipole symmetry, the quadrupole modes are also strongly localized inside the cavity shell.
 
\begin{figure}
\centering
\includegraphics[]{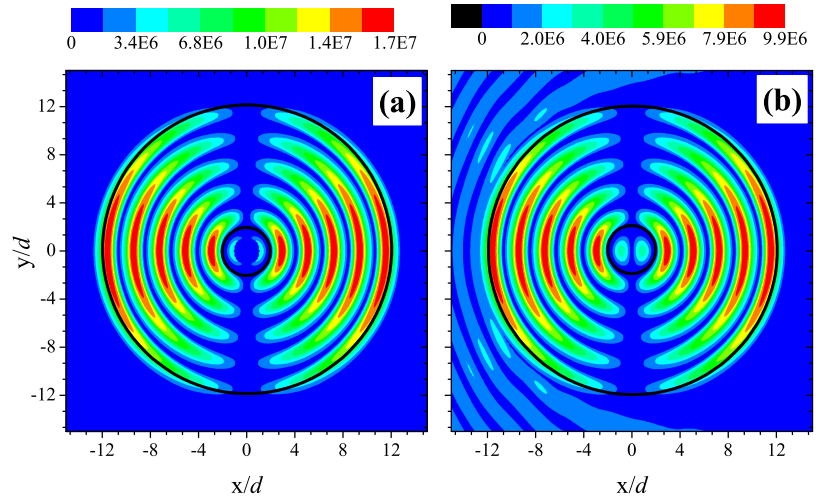} 
\caption{\label{fig:modefabryq1}Pressure maps (amplitude) of the Fabry-Perot mode of dipole symmetry ($q=$1) that appears at frequency 0.3648 (see Fig. \ref{fig:Zoom}. 
(a) Excitation by a punctual sound source (not depicted) put inside the central cavity; at $(x,y)=(0,d)$. (b) Excitation by external sound with a plane wavefront. 
The black circles define the borders of the RSC shell.}
\end{figure}

\begin{figure}
\centering
\includegraphics[]{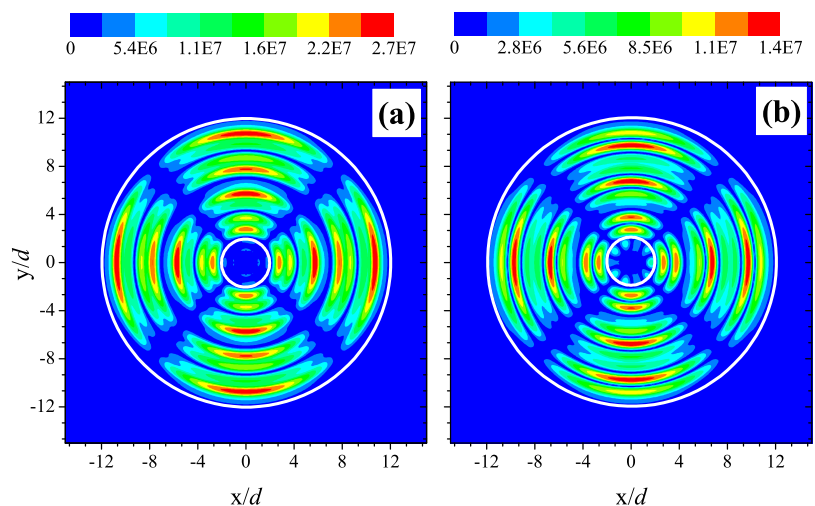} 
\caption{\label{fig:modefabryq2in}Pressure maps (amplitude) of the first two Fabry-Perot like modes existing in the RSC shell with quadrupole symmetry ($q=$2). 
Their frequencies are 0.5936 (a) and 0.6135 (b), respectively. They are excited by a punctual sound source (not depicted) put inside the central cavity; at $(x,y)=(0,d)$. 
The white circles define the borders of the RSC shell.}
\end{figure}

\begin{figure}
\centering
\includegraphics[]{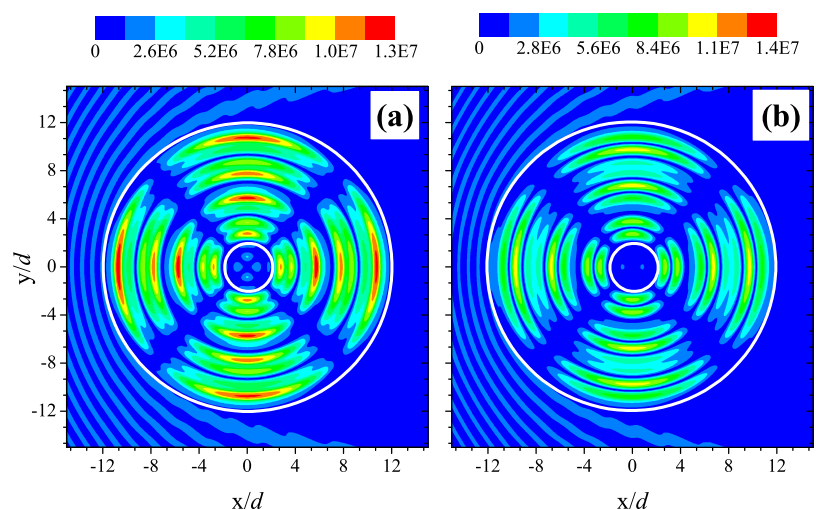}  
\caption{\label{fig:modefabryq2ext}Pressure maps (amplitude) calculated for the scattering of the RSC shell by an external sound with a plane wavefront impinging from the left side of the $x-$axis. The impinging sound has the same frequency than Fabry-Perot modes characterized in Fig. \ref{fig:modefabryq2in}:(a) 0.5936 and (b) 0.6135, respectively.}
\end{figure}

 Finally, Fabry-Perot like mode with dipole and quadrupole symmetry are shown in Figs. \ref{fig:modefabryq1}, \ref{fig:modefabryq2in} and \ref{fig:modefabryq2ext} under the two different excitation conditions. Note that these modes are contained in the RSC shell, which in all the cases have 10 layers. Also note in Fig. \ref{fig:modefabryq1}(b) that the dipole is excited along the direction of the impinging wave, the $x-$axis. This property could be used as a sonic device to determine the direction of external sound sources and even to determine its exact positions by using an additional RSC shell\cite{torrentPRL09}. In regards with quadrupole modes, a feature to remark is related to the additional number of oscillations that present the mode with frequency 0.6135 in comparison with that having lower frequency (0.5936). This difference is related with the higher number of oscillations that the periodic functions with larger argument $K_q(R_a-R_b)$ have.
    

%
%
\section{\label{sec:summary}Summary}
We have studied the resonances existing in two-dimensional (2D) RSC shells, which are acoustic structures consisting of a circular cavity surrounded by a finite RSC, both being embedded in a fluid background. Two type of resonant modes have been clearly characterized. On the one hand, cavity like modes, which have frequencies within the acoustic bandgaps of the RSC and are characterized by their strong localization inside the cavity. On the other hand, Fabry-Perot like modes, which appear in the frequency regions where the RSC have an acoustic passband. The later modes are equivalent to those previously characterized in sonic crystal slabs\cite{sanchisPRB03} and are confined in the shell; i.e., between the cavity and the embedded background. 

The main advantage of the RSC shells is the possibility of selecting the mode symmetry either inside the cavity as well as in the shell by using bandgap engineering of the built in RSC. This mode symmetry selection can be used in designing devices for beam forming of the radiation field. Moreover, since the Fabry-Perot resonances are selectively excited along the direction of the external sound source, the dipole modes could be used in acoustic devices capable to determine the exact position of external sound sources. Moreover, RSC shells are excellent candidates for designing devices with functionality similar to the cochlea by following a recent proposal by Bazan and coworkers\cite{bazanAPL07}, who used circular cavities with such purpose. In this regards, RSC shells have the advantage of their large tailoring possibilities. 

Let us remark that results obtained for 2D shells can be extrapolated to their three-dimensional (3D) counterparts, in which the cavity defect is a spherical hole. The high $Q$ values that these structures present, either in 2D as well in 3D, make them suitable to fabricate acoustic resonators with possible applications in sensing and metrology\cite{miklosRSI01}.

With a few modifications, the results described here can be extended in the realm of electromagnetic waves, where radial photonic crystals (RPC) have been also proposed\cite{torrentPRL09}. In this context, photonic cavities based on RPC shells are expected to have modes with huge $Q-$factors, a property that can be used to fabricate low threshold microcavity lasers with selected symmetry.

\ack
We thank the financial support provided by the USA Office of Naval Research (Grant No. N000140910554) and the Spanish Ministry of Science and Innovation under Contracts No. TEC2007-67239 and No. CSD2008-00066 (CONSOLIDER Program).

\appendix
\section{Demonstration of $\left|M_s\right|=$1}
Let us define the matrix $\hat{M}_{s}(r)$ as
\begin{equation}
 \hat{M}_s(r)=\left(\begin{array}{cc}\phi_q^+(r) & \phi_q^-(r)\\ r\rho_r^{-1}(r)\partial_r\phi_q^+(r) & r\rho_r^{-1}(r)\partial_r\phi_q^-(r)\end{array}\right)
\end{equation} 
The matrix $M_s$ can be write as
\begin{equation}
 M_s=\hat{M}_s^{-1}(R_b)\hat{M}_s(R_a),
\end{equation} 
and the determinant of $M_s$ is 
\begin{equation}
 |M_s|=\frac{|\hat{M}_s(R_a)|}{|\hat{M}_s(R_b)|}
\end{equation} 
Applying the equation \eref{eq:waveqr} it can be shown that the derivative of determinant of $\left|\hat{M}_s(r)\right|$ is zero, so that this quantity is not a function of $r$, that is,
\begin{eqnarray}
 \partial_r|\hat{M}_s(r)|=&\nonumber\\\partial_r\left[\phi_q^+(r) r\rho_r^{-1}(r)\partial_r\phi_q^-(r)\right]-&\partial_r\left[\phi_q^-(r) r\rho_r^{-1}(r)\partial_r\phi_q^+(r)\right]=0
\end{eqnarray} 
so that 
\begin{equation}
 |M_s|=1
\end{equation} 
\section*{References}

\end{document}